\documentclass[prl,aps,floats,twocolumn, preprintnumbers,tightenlines]{revtex4}
\usepackage{epsfig}
\usepackage{amsmath}
\usepackage{amssymb}
\usepackage{amsthm}
\usepackage{graphics}
\usepackage{epsfig}
\newcommand{\phib}{\mbox{\boldmath$\varphi$}}

\newcommand{\bea}{\begin{eqnarray}}

\newcommand{\eea}{\end{eqnarray}}
\newcommand{\beq}{\begin{equation}}
\newcommand{\eeq}{\end{equation}}
\begin{document}

\title{Dynamics of domain walls intersecting black holes}

\author{Antonino Flachi${}^1$%
\footnote{flachi@yukawa.kyoto-u.ac.jp}, %
Oriol Pujol\`as${}^{1,2}$%
\footnote{pujolas@ccpp.nyu.edu}, %
Misao Sasaki${}^1$%
\footnote{misao@yukawa.kyoto-u.ac.jp} and %
Takahiro Tanaka${}^3$%
\footnote{tama@scphys.kyoto-u.ac.jp}}%
\address{$^{1}$Yukawa Institute for Theoretical Physics, Kyoto
University, Kyoto 606-8503, Japan\\
$^{2}$Center for Cosmology and
Particle Physics, Department of Physics, New York University, 4
Washington Place, New York, NY
10003 US\\
$^{3}$Department of Physics, Kyoto University, Kyoto
606-8502, Japan} 
\preprint{YITP-05-71}
\preprint{KUNS-2008} 

\begin{abstract}
Previous studies concerning the interaction of branes and black holes suggested that a small black hole intersecting a brane may escape via a mechanism of reconnection.
Here we consider this problem by studying the interaction of a small
 black hole and a domain wall composed of a scalar field and simulate the evolution of this system when the black hole acquires an initial recoil velocity.
We test and confirm previous results, however, unlike the cases previously studied, in the more general set-up considered here, we are able to follow the evolution of the system also during the separation, and completely illustrate how the escape of the black hole takes place. 
\end{abstract}
\maketitle

\noindent

{\it Introduction.}  It is well known that primordial black holes and
domain walls may have formed in the early universe: on the one hand,
cosmological density perturbations may have collapsed and formed small
black holes~\cite{pbh1,pbh2}, on the other hand, during the cooling phase after the big bang, the series of phase transitions that have occurred may have produced extended topological structures like domain walls~\cite{book}. 
In view of the stringent constraints that both primordial black holes and domain walls provide,  
understanding how these objects interact could help in clarifying many issues regarding the early universe and, possibly, be of phenomenological relevance.

In the past few years, the interaction between black holes and domain
walls has been object of some study. Of particular significance, in
relation to our work, are the results of
Refs.~\cite{frolov1,dm1,dm2,dm3,dm4}. Ref.~\cite{frolov1} considers the
Dirac-Nambu-Goto approximation for the domain wall and proves the
existence of a family of static wall solutions intersecting the
black hole event horizon. Subsequent work extended those results to the
case of thick walls, Refs.~\cite{dm1,dm2,dm3,dm4}.

An important problem where the previous studies may find application is
the scattering of a black hole and a domain wall~\cite{stoj}.  When the
domain wall moves towards the black hole, it is expected to be captured
by the black hole and the static configurations mentioned above may
describe this process in the adiabatic limit, namely when the relative
motion is very slow. It is likely that the membrane will experience
large deformations and topology changing processes will take place, but
the adiabatic limit cannot easily resolve this issue.
A full dynamical simulation is necessary, and it would also help to
clarify what happens in more extreme situations, which are more likely
to have occurred in the early universe, where the adiabatic
approximation may not be adequate. 

Aside from the motivations mentioned above, the exciting
possibility of observing mini black holes at forthcoming collider
experiments, as predicted by low scale gravity
theories~\cite{add,rs,dam,dl,gt}, has rejuvenated interest in the
subject. 
In such models our universe has a domain wall structure, {\it the brane}, that is embedded in a higher dimensional space; the standard model
particles are confined on the brane and gravity propagates throughout the higher dimensional bulk spacetime. Many
realizations of this scenario result in lowering the Planck scale to a
few TeV and share the common prediction that a small black hole forms
when two particles collide at sufficiently high energy with small impact
parameter. Once the 
black hole is produced, it will emit Hawking radiation, and this can in
principle be observed at the CERN LHC or in cosmic ray facilities. 
The radiation of the black hole will partly go into lower dimensional
fields, localised on the brane, and partly into higher dimensional
modes, which will cause the black hole to recoil into the extra
dimensions. 

The recoil of the black hole in the context described above was first
studied in Ref.~\cite{frolov} within a toy model consisting of two
scalar fields, one describing the black hole and the other a possible
quanta emitted by the black hole in the process of evaporation; the
coupling between them is fixed as to reproduce the probability of
emission of a scalar quanta from the black hole as prescribed by the
Hawking formula, and a delta function potential is used to mimic the
interaction between the black hole and the brane. In the approximation
that the interaction with the brane is negligibly small, it is shown
that, as soon as a quanta is emitted in the extra dimension, the black
hole will slide off the brane. Although this has the important
observable effect of a sudden interruption of the radiation on the brane
(Ref.~\cite{stojkovic} explores some possible phenomenological
consequences), it is not clear how the separation process occurs.
To clarify this issue and to have further evidence for the conclusions of
Ref.~\cite{frolov}, two of the present authors have considered the interaction between
`mini' black holes and branes from the different perspective of studying
the dynamics of branes in black hole spacetimes,
Ref.\cite{flachi}. Specifically, by treating the brane in the
Dirac-Nambu-Goto approximation, we have shown that, once the black hole
acquires an initial recoil velocity perpendicularly to the brane, an
instability develops and the brane tends to envelop the black hole. This
suggests a mechanism of escape for the black hole due to the
reconnection of the brane. In some approximation the time of escape can
be estimated and it was found to be shorter than the evaporation
time. This fact may have important phenomenological consequences.

The previous claim certainly deserves reconsideration and one of our
motivations is to test the robustness of the conclusions of
Ref.~\cite{flachi} by modeling the brane as a domain wall composed
of a scalar field. Also, unlike
the Dirac-Nambu-Goto case previously studied, modeling the brane as a
domain wall allows one to describe reconnection phenomena. In
our case, this means that we can follow the evolution of the system
while the separation of the black hole occurs, and completely illustrate
how the escape takes place.

{\it Domain wall dynamics.} The system we intend to study consists of a domain wall intersecting a black hole.
As customary, we consider black holes whose mass $m$
exceeds the fundamental Planck scale $M$, in order to
make quantum gravity corrections negligible. Also, the size of the
mini black hole is assumed to be smaller than the characteristic
length of the extra dimensions. We shall restrict our analysis to the
case when the wall tension is small enough, as this allows us to
ignore the self-gravity of the wall.
In this regime, the mini black hole is completely immersed in the
higher dimensional spacetime and can be adequately described
by asymptotically flat solutions~\cite{tangherlini,mp}
\beq %
ds^2 = -f(r) dt^2 + f^{-1}(r) dr^2 +r^2d\Omega_{d-2}^2~,
\label{eq1} %
\eeq%
where $d\Omega_{d-2}$ is the line element of a
$(d-2)-$dimensional unit sphere,
$$
f(r)= 1-\left({r_H\over  r}\right)^{d-3}
$$
and $r_H$ is the horizon radius. We use units in which
$c=\hbar=1$.\\

Since the character of vacuum structures is generally insensitive
to the details of the model, we limit ourselves to a scalar
effective field theory with quartic potential
\beq%
\label{pot}%
V(\Phi) = {\lambda\over 4}(\Phi^2 -\eta^2)^2~.%
\eeq%
The domain wall solutions of this model in flat space have tension
$\sigma\simeq\sqrt\lambda\eta^3$ and thickness
${\tt t} \simeq1/\sqrt\lambda\eta$.
The coupling constant $\lambda$ is of order $\Lambda^{4-d}$, with
$\Lambda\lesssim M$ being the cutoff of the effective field 
theory. We can neglect 
the quantum corrections to the domain wall-type configurations 
at weak coupling $\Lambda^{(d-2)/2}\gg\eta$. 
This implies $\sigma\gg {\tt t}^{1-d}$, and hence we have 
$\sigma r_H^{d-2}/M\gg (r_H/{\tt t})^{d-1}(M_d/M)^{(d-2)/(d-3)}$, 
where we introduced the $d$-dimensional Planck mass $M_d=G^{1/(d-2)}$. 
Requirement of negligible self-gravity therefore implies that 
the wall is not very thin and/or 
the mass of the black hole is sufficiently large. 

The evolution equations describing the dynamics of the domain wall
intersecting a black hole, are given by
\bea%
0 &=& -f^{-1}(r) \partial_t^2 \Phi + {1\over r^{d-2}} \partial_r
\left( r^{d-2} f(r) \partial_r \Phi \right)
\nonumber \\%
&+& {1\over r^2 (\sin\theta)^{d-3} } \partial_{\theta} \left[
(\sin\theta)^{d-3} \partial_{\theta} \Phi \right] -  V'(\Phi) ~,
\label{eq9}%
\eea%
where we assumed $O(d-2)$-rotational symmetry around 
the axis perpendicular to the equatorial plane of the black hole.

The initial conditions are specified so as to 
mimic the initial recoil of the black hole.
For this purpose, we impose that 
$\Phi$ and $\partial \Phi$ at $t=0$ are  
given by a static kink profile in the flat spacetime 
boosted in the direction of the symmetry axis
\beq%
\Phi_v (t,r,\theta) = \eta \tanh \left(%
\sqrt{\lambda\over 2}~ \eta ~%
{r\cos \theta - vt\over \sqrt{1-v^2}}  \right)~,%
\label{eq15}%
\eeq%
with $v$ constant.

Boundary conditions have to be specified on the
outer boundary, on the symmetry axis and on the horizon. At `infinity', since the gravity
of the black hole is expected to be negligible, we assume that
$\Phi$ reduces to the flat boosted form (\ref{eq15}). The regularity conditions at the symmetry axis 
and on the horizon do require
\begin{eqnarray}%
\partial_\theta \Phi(t,r,\theta) & = &0~,~~~\mbox{at} ~\theta=0,~\pi~,%
\label{eq10}%
\cr
\partial_t \Phi(t, r,\theta) &=& 0~,~~~\mbox{at} ~r=r_H~.%
\label{eq11}%
\end{eqnarray}%

In order to solve Eq.~(\ref{eq9}) numerically, we use
a mixed spectral and finite difference method. The first step is to decompose the
general solution in terms of a complete basis of smooth global
functions~\cite{spectral}. Here, for convenience, we adopt the
Chebyshev polynomials of second kind, and write the general
solution as
\beq%
\Phi(t,r,\theta) = \sum_{n=0}^{N} \varphi_n(t,r) U_n(\cos
\theta)~, \label{expaT}
\eeq%
where $N$ is the number of harmonics used. The boundary condition
along the symmetry axis is trivially satisfied, as the Chebyshev
polynomials $U_n(\cos \theta)$ are regular at $\theta=0,\pi$. The
decomposition (\ref{expaT}) is used as ans$\ddot{\mbox{a}}$tze in
equation (\ref{eq9}), and some trivial algebra allows us to recast
Eq.~(\ref{eq9}) in matrix form
\bea%
\ddot{\phib} &=& f^2(r) \phib'' + f(r) \left[ f'(r) + {d-2\over r} f(r) \right] \phib' \nonumber \\
&+&{f(r) \over r^{2}} \ {\mathbb L} \cdot \phib  
+ {\bf F}~,
\label{system}
\eea%
where we have defined
\bea%
{\mathbb L}_{nm} &=& {2\over \pi} \int_{-1}^{1}
{\partial_x \left[(1-x^2)^{d-2\over 2}\partial_x U_n \right]U_m \over (1-x^2)^{{d-1\over 2}-2} } dx,\\
{\bf F}&=& -{2f(r)\over \pi}\int_{-1}^{1}   {\bf
U}(x)V'(\Phi)\sqrt{1-x^2} dx~,
\label{FF}
\eea%
$x=\cos\theta$, and the vectors $\phib$  and ${\bf U}$ are given
by
\bea%
\phib &=& (\varphi_0(t,r), \varphi_1(t,r), \cdots, \varphi_{N}(t,r))~,\nonumber \\
{\bf U} &=& (U_0(x), U_1(x), \cdots, U_{N}(x))~.\nonumber
\eea%
Equation (\ref{system}) is a system of coupled two-dimensional
non-linear partial differential equations, which can be solved by
using finite difference methods. 

We define 
$\phib^-$, $\phib$ and $\phib^+$, which represent the
configurations at three subsequent time steps, $t_-=t
-\delta_t$, $t$ and $t_+=t +\delta_t$, respectively. We now define
a one-dimensional grid along the radial direction, and label the
points as $r_{i}$. Then the values of a function $g(t,r)$ on a grid point $r_i$ at $t=t_-, t$ and $t_+$ will be denoted by
$g^-_{i}$, $g_i$ and $g^+_{i}$.
In terms of the coefficients of the
expansion, the initial conditions are written as
\beq%
\phib^-_{i} = {2\over \pi} \int_{-1}^{1} \Phi_v(0, r_i, x) {\bf
U}(x)\sqrt{1-x^2}d{x}~,
\eeq%
and
\beq%
\phib_i = {2\over \pi} \int_{-1}^{1} \Phi_{v}(\delta_t, r_i,x)
{\bf U}(x) \sqrt{1-x^2}d{x}~,
\eeq%
where $\Phi_v(t,r,x)$ is the boosted domain wall solution, 
Eq.~(\ref{eq15}).
As for the boundary conditions, we have
\beq%
\phib(t,r \geq r_\infty) = {2\over \pi} \int_{-1}^{1} \Phi_v (t,r
\geq r_\infty,x) {\bf U}(x)\sqrt{1-x^2}d{x}~.\nonumber
\eeq%
at the outer boundary $r_\infty$, and
\beq%
\dot{\phib}(t,r=r_H) = 0~,\nonumber
\eeq%
on the horizon. 

To solve the system of Eqs.~(\ref{system}) we use a central difference discretization scheme. In this way the R.H.S. of (\ref{system}) can solely be expressed in terms of $\phib$, whereas the L.H.S. will depend on $\phib^-$ and $\phib^+$. Therefore, once the boundary and initial conditions are specified, the system of equations (\ref{system}) is completely determined. The potential term ${\bf F}$ is evaluated by first reconstructing the field profile $\Phi$ and then use it to evaluate the function $V'(\Phi)$. As last step, $V'(\Phi)$ is convoluted with the basis functions according to formula (\ref{FF}).

\begin{figure}
\scalebox{0.5} {\includegraphics{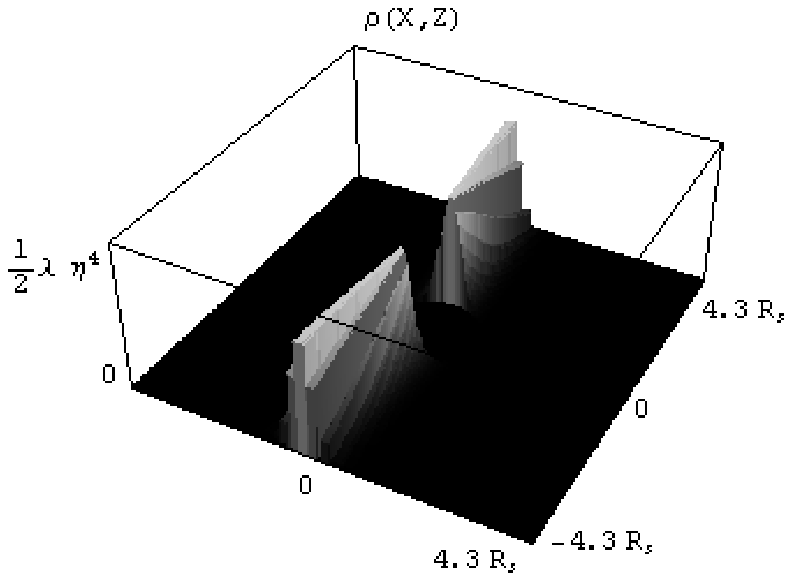}}
\scalebox{0.5} {\includegraphics{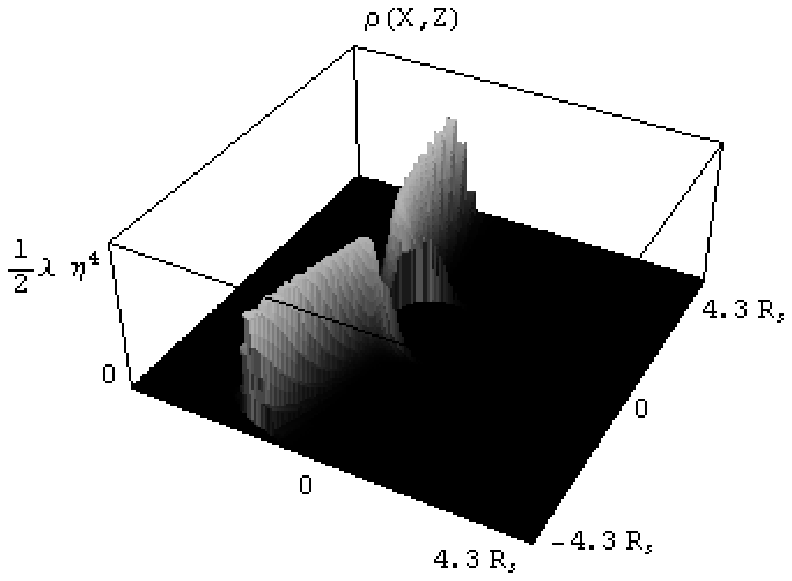}}
\scalebox{0.5} {\includegraphics{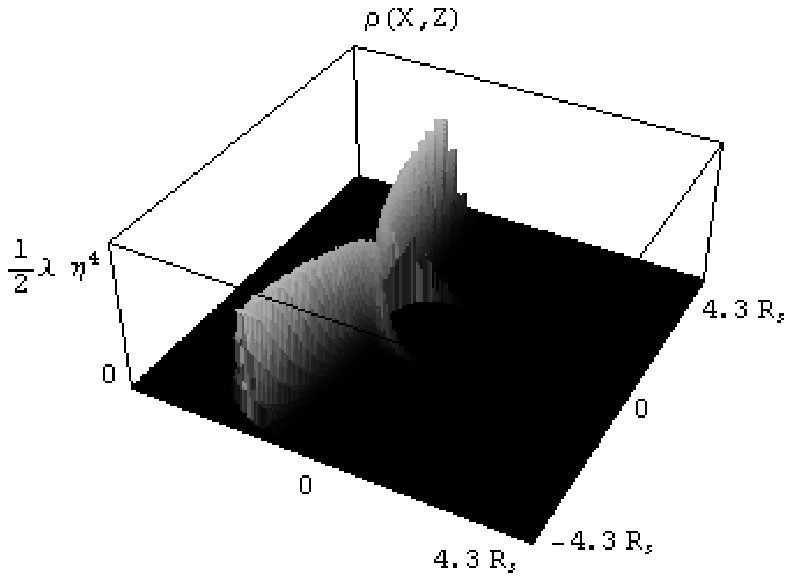}}
\scalebox{0.5} {\includegraphics{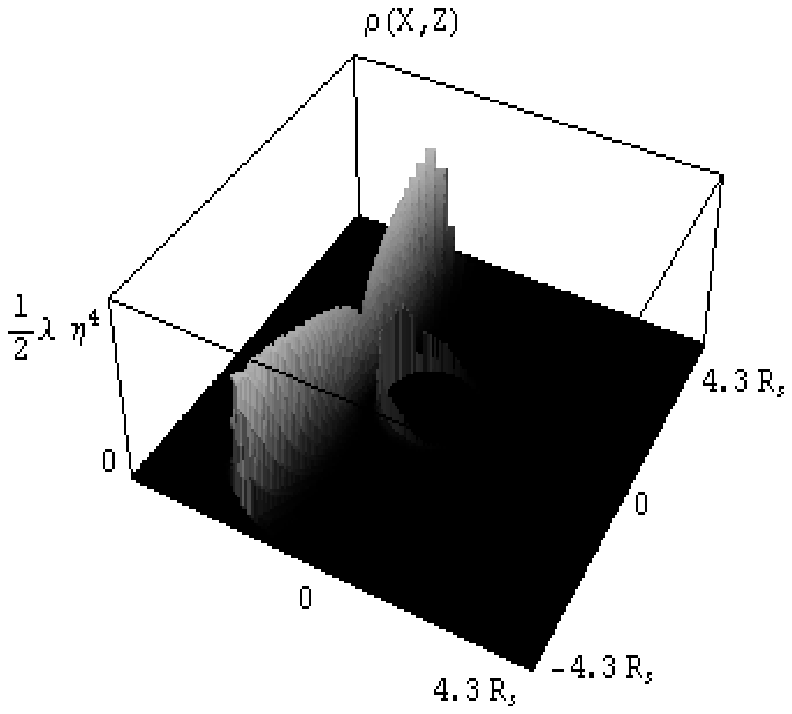}}
\scalebox{0.5} {\includegraphics{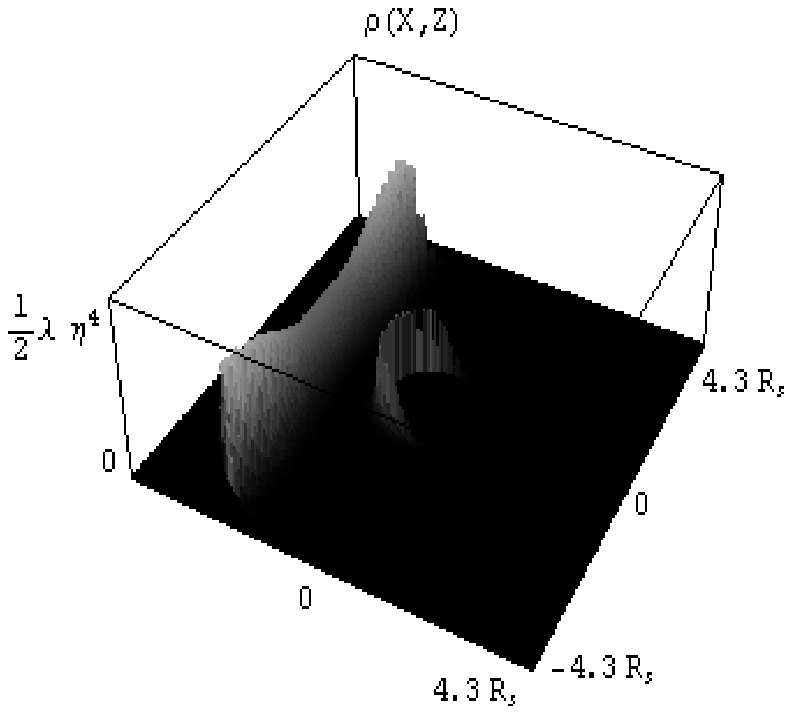}}
\scalebox{0.5} {\includegraphics{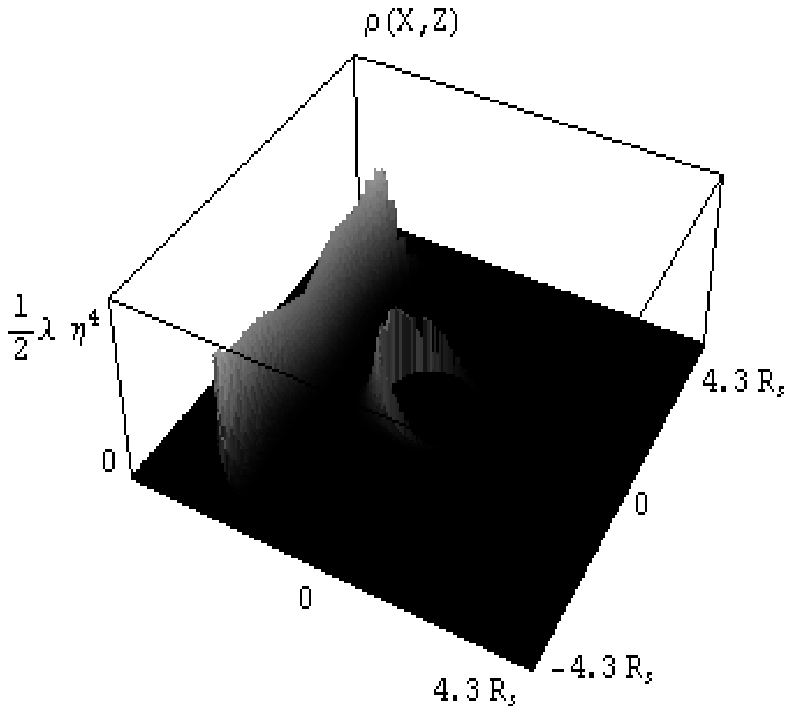}}
\caption{
Evolution of the domain wall. The figure shows how the energy density of
 the wall evolves, explicitly illustrating the separation process and
 escape of the black hole. The reported simulation refers to the
 five dimensional case, $d=5$, 
 and a domain wall whose thickness is
 $1/3$ of the Schwarzschild radius and $v=0.3$.}
\label{simulation}
\end{figure}

The details of the simulations will be reported elsewhere, here we briefly illustrate the main result.
Fig.~\ref{simulation} shows six snapshots describing the evolution of the energy density of the domain wall,
\beq%
\rho(t,r,\theta) = {1\over 2f(r)} \Phi_{,t}^2 +{f^{}(r)\over 2}
\Phi_{,r}^2+{1\over 2r^2} \Phi_{,\theta}^2 + V(\Phi)~.
\eeq%
The plot presents the result of a simulation obtained by using  $N=30$
harmonics and {100 grid points} in the radial direction.  Standard
numerical checks have been performed and the evolution has been carried
out in various cases with the thickness of the domain wall ranging from
$0.01$ to $1$ times the size of the horizon radius and for various
initial velocities, $v$ ranging from $0.01$ to $1$.
The results obtained in the case of a domain wall confirm the suggestion
of Ref.~\cite{frolov,flachi}: the domain wall envelops the black hole,
that completely separates from the mother brane; after the pinching
occurs, the mother brane relaxes into a stable configuration.
 
Modeling the brane as a domain wall, we have given further evidence in
support of the conclusions of Refs.~\cite{frolov,stojkovic, flachi} that the black hole intersecting a brane may escape once
it acquires an initial recoil velocity. As in the previous work, we have
neglected the self-gravity of the wall, but the present treatment
allows us to describe reconnection processes of the domain wall. By
following the evolution of the system during the separation, we
succeeded in illustrating completely how this process takes place.

There is one important question which has not been considered here:
what happens when the tension is non-negligible?
Intuition suggests that the back reaction may prevent the separation at
least for small initial velocities, 
and an energy barrier has to be overcome to
induce instability leading to pinching. In other words, when the tension
of the brane is switched on, it might be possible to find a sequence of
static configurations before we encounter a critical unstable
configuration above which the instability sets in. However this problem
is rather complicated and beyond the scope of the present article. Study to
understand this issue is in progress and we hope to report on this soon.

{\it Acknowledgements}. We thank J.J. Blanco-Pillado, H. Kudoh,
T. Shiromizu and H. Yoshino for useful discussions and for raising
interesting questions. OP acknowledges the kind hospitality of the CCPP
at NYU. This work is supported in part by Grant-in-Aid for Scientific
Research, Nos. 1604724 and 16740165 and 
the Japan-U.K. Research Cooperative Program
both from Japan Society for Promotion of Science.  
This work is also supported by the 21st Century COE ``Center for
Diversity and Universality in Physics'' at Kyoto university, 
from the Ministry of Education, Culture, Sports, Science and 
Technology of Japan and by Monbukagakusho Grant-in-Aid for
Scientific Research(S) No. 14102004 and (B) No.~17340075.
A.F. is supported by the JSPS under contract No. P047724. O.P. is supported by
the JSPS under contract No. P03193.

\end{document}